\documentclass[twocolumn,aps,amssymb,floatfix,superscriptaddress,preprintnumbers]{revtex4}
\usepackage[usenames]{color}

\usepackage{graphicx}
\usepackage{bm}
\usepackage{amsmath}
\usepackage{amssymb}
\usepackage{amsfonts}
\usepackage{float}
\usepackage{dsfont}  
\usepackage{slashed}  
\usepackage{booktabs}
\usepackage{multirow}
\usepackage{subfigure}
\usepackage{soul}
\usepackage[sort&compress]{natbib}
\usepackage[normalem]{ulem}
\usepackage{adjustbox}

\usepackage{hyperref}
\usepackage{xr-hyper}
\makeatletter
\newcommand*{\addFileDependency}[1]{
  \typeout{(#1)}
  \@addtofilelist{#1}
  \IfFileExists{#1}{}{\typeout{No file #1.}}
}
\makeatother

\newcommand{\be}{\begin{equation}}  
\newcommand{\ee}{\end{equation}}  
\newcommand{\beq}{\begin{eqnarray}}  
\newcommand{\eeq}{\end{eqnarray}}

\newcommand{\bea}{\begin{eqnarray}}
\newcommand{\eea}{\end{eqnarray}}

\newcommand{\UCY}{Department of Physics, University of Cyprus, P.O. Box 20537, 1678 Nicosia, Cyprus}
\newcommand{\CYI}{  Computation-based Science and Technology Research Center, The Cyprus Institute, 20 Kavafi Street, Nicosia 2121, Cyprus}
\newcommand{\TU}{Department of Physics, Temple University, Philadelphia, PA 19122 - 1801, USA }
\newcommand{\NIC}{NIC, DESY Zeuthen, Germany}
\newcommand{\HPCA}{High Performance Computing and Analytics Lab, Rheinische Friedrich-Wilhelms-Universit\"at Bonn,\\ Friedrich-Hirzebruch-Allee 8, 53115 Bonn, Germany}
\newcommand{\HI}{Helmholtz-Institut für Strahlen- und Kernphysik, University of Bonn, Germany}
\newcommand{\BCTP}{Bethe Center for Theoretical Physics, University of Bonn, Germany}
\newcommand{\INFN}{Dipartimento di Fisica and INFN, Universit\`a di Roma ``Tor Vergata",\\Via della Ricerca Scientifica 1, I-00133 Roma, Italy}
\newcommand{\UDP}{Dipartimento di Scienze Matematiche, Fisiche e Informatiche,
 Universit\`a di Parma and INFN,  Gruppo  Collegato di Parma
Parco Area delle Scienze 7/a (Campus), 43124 Parma, Italy}

\begin{document}

\title{First moments of the nucleon transverse quark spin densities using lattice QCD}

\author{C.~Alexandrou}\affiliation{\UCY}\affiliation{\CYI}
\author{S.~Bacchio}\affiliation{\CYI}
\author{M.~Constantinou}\affiliation{\TU}
\author{P.~Dimopoulos}\affiliation{\UDP}
\author{J.~Finkenrath}\affiliation{\CYI}
\author{R.~Frezzotti}\affiliation{\INFN}
\author{K.~Hadjiyiannakou}\affiliation{\UCY}\affiliation{\CYI}
\author{K.~Jansen}\affiliation{\NIC}
\author{B.~Kostrzewa}\affiliation{\HPCA}
\author{G.~Koutsou}\affiliation{\CYI}
\author{G.~Spanoudes}\affiliation{\CYI}
\author{C.~Urbach}\affiliation{\BCTP}\affiliation{\HI}

\begin{abstract}
\noindent We present a calculation of the Mellin moments of the transverse quark spin densities in the nucleon using lattice QCD. The densities are extracted from the unpolarized and transversity generalized form factors extrapolated to the continuum limit using three $N_f=2+1+1$ twisted mass fermion gauge ensembles simulated with physical quark masses and spanning three lattice spacings. The first moment of transversely polarized quarks in an unpolarized nucleon shows an interesting distortion, which can be traced back to the sharp falloff of the transversity generalized form factor $\bar{B}_{Tn0}(t)$. The isovector tensor anomalous magnetic moment is determined to be $\kappa_T=1.051(94)$, which confirms a negative and large Boer-Mulders function, $h_1^{\perp}$, in the nucleon.  
 
\end{abstract}

\maketitle

\noindent
\textit{Introduction:}
\noindent
Understanding the spin content of the nucleon is of paramount importance for hadron structure. While significant progress has been made in recent years revealing the longitudinal spin structure of the nucleon~\cite{Ji:2020ena,Alexandrou:2020sml,Wang:2021vqy}, the transverse spin structure remains lesser known from phenomenology~\cite{Radici:2015mwa,Kang:2015msa,Radici:2018iag}, a situation that will improve with results from planned experiments (SoLID~\cite{Chen:2014psa,Zhao:2019yod}, Electron-Ion Collider~\cite{AbdulKhalek:2021gbh}). In lattice QCD, theoretical progress~\cite{Ji:2013dva,Radyushkin:2017cyf} has enabled the extraction of the x-dependence of parton distribution functions (PDFs) at the physical pion mass~\cite{Alexandrou:2018eet,Alexandrou:2021bbo,Alexandrou:2021oih,Alexandrou:2019lfo,Joo:2020spy,Bhat:2020ktg}, as well as first results on generalized parton distributions (GPDs)~\cite{Alexandrou:2021bbo}. For a summary of these approaches we refer the reader to~\cite{Cichy:2018mum,Ji:2020ect,Constantinou:2020pek,Constantinou:2020hdm,Cichy:2021lih}.

In this work, we use lattice QCD for the study of the transverse spin properties of the nucleon by considering the first two Mellin moments of the 3-dimensional (3D) probability densities $\rho(x,\bold{b}_\perp,\bold{s}_\perp,\bold{S}_\perp)$, where $x$ is the longitudinal momentum fraction, $\bold{s}_\perp$ the transverse quark spin, $\bold{b}_\perp$ the transverse vector from the center of momentum of the nucleon, and $\bold{S}_\perp$ the transverse spin of the nucleon. As discussed in Ref.~\cite{Diehl:2005jf}, to access the transverse spin densities one needs to compute the twist-two matrix elements of the chiral-even unpolarized and chiral-odd transversity GPDs. The probability density~\cite{Diehl:2005jf} is then given as
\begin{eqnarray}
  &&\rho(x,\bold{b}_\perp,\bold{s}_\perp,\bold{S}_\perp) = \frac{1}{2} \bigg[ H(x,b^2_\perp) + \nonumber \\ &&
       \frac{\bold{b}_\perp^j \epsilon^{ji}}{m_N} \left( \bold{S}_\perp^i E'(x,b^2_\perp) + \bold{s}_\perp^i \bar{E}'_{T}(x,b^2_\perp)\right) + \nonumber \\ &&
           \bold{s}_\perp^i \bold{S}_\perp^i \left( H_T(x,b^2_\perp) - \frac{\Delta_{b_\perp} \tilde{H}_T(x,b^2_\perp)}{4 m_N^2} \right) +
            \nonumber \\ &&
     \bold{s}_\perp^i(2\bold{b}_\perp^i \bold{b}_\perp^j- \delta^{ij} b^2_\perp)\bold{S}_\perp^j \frac{\tilde{H}''_T(x,b^2_\perp)}{m_N^2} \bigg].
     \label{Eq:rho}
\end{eqnarray}
The GPDs $H,\,E,\,H_T,\,E_T,\,\widetilde{H}_T$ involved in Eq.~\eqref{Eq:rho} are given in the impact parameter space for zero skewness by a Fourier transformation, $\bold{\Delta}_\perp \leftrightarrow \bold{b}_\perp$, where $\bold{\Delta}_\perp$ is the transverse momentum transfer and $-t\equiv \Delta^2$. $m_N$ is the nucleon mass, $\epsilon^{ij}$ is the antisymmetric tensor and the derivatives are denoted as $F' \equiv \frac{\partial F}{\partial b^2}$ and $\Delta_{b_\perp} F \equiv 4 \frac{\partial}{\partial b^2} (b^2_\perp \frac{\partial}{\partial b^2_\perp}) F$. The GPD $\bar{E}_T$ is defined as a linear combination of two GPDs, namely $\bar{E}_T \equiv E_T + 2 \tilde{H}_T$. The moments are then computed as an integral over the momentum fraction as
\begin{equation}
    \langle x^{n-1} \rangle_\rho (\bold{b}_\perp,\bold{s}_\perp,\bold{S}_\perp) \equiv \int_{-1}^{1} dx ~ x^{n-1} \rho(x,\bold{b}_\perp,\bold{s}_\perp,\bold{S}_\perp),
\end{equation}
where $n$ is a positive non-zero integer corresonding to the $n$\textsuperscript{th}-moment. The GPDs reduce to the generalized form factors (GFFs) if integrated over $x$. For the unpolarized case, we have $A_{n0} = \int dx~x^{n-1} H$ , $B_{n0} = \int dx~x^{n-1} E$ and $\tilde{A}_{n0} = \int dx~x^{n-1} \tilde{H}$, for zero skewness, and analogously  for the tensor GFFs. 

In this work, we are interested in GFFs that parameterize off-forward nucleon matrix elements of local vector and tensor quark operators, defined as
\begin{equation}
    {\cal O}^{\mu}_V = \bar{q}(x) \gamma^\mu q(x), \quad {\cal O}^{\mu\nu}_{VD} = \bar{q}(x) \gamma^{\{\mu} i \overleftrightarrow{D}^{\nu\}} q(x),
    \label{Eq:vector}
\end{equation}
\begin{equation}
    {\cal O}^{\mu\nu}_T = \bar{q}(x) \sigma^{\mu\nu} q(x),\;\;
    {\cal O}^{\mu\nu\rho}_{TD} = \bar{q}(x) \sigma^{[\mu\{\nu]} i \overleftrightarrow{D}^{\rho\}} q(x),
    \label{Eq:tensor}
\end{equation}
where $\overleftrightarrow{D}$ is the symmetrized covariant derivative, $\{\cdots\}$ denotes symmetrization and subtraction of the trace and  $[\cdots]$ antisymmetrization of the enclosed indices. For details on how the nucleon matrix elements of the operators in Eqs.~(\ref{Eq:vector}) and~(\ref{Eq:tensor}) yield the GFFs we refer to Ref.~\cite{Diehl:2003ny}.

\noindent
\textit{Lattice methodology:}
\noindent 
We employ the twisted-mass fermion discretization scheme~\cite{Frezzotti:2003ni,Frezzotti:2000nk}, which provides automatic ${\cal O}(a)$-improvement for both physical observables and renormalization constants~\cite{ETM:2010iwh}. The ensembles are generated with two mass-degenerate light, a strange, and a charm quark, referred to as $N_f=2+1+1$. The bare light quark mass is tuned to reproduce the isosymmetric pion mass $m_\pi=0.135$~MeV within 1-4 MeV while the heavy quark masses are tuned with inputs given by the physical kaon and D-meson masses as well as the D-meson
decay constant, following the procedure of Refs.~\cite{Finkenrath:2022eon,Alexandrou:2018egz}. The parameters of the ensembles analyzed in this work can be found in Table~\ref{tab:ens}. We note that the lattice spacing has been determined from the nucleon mass as discussed in Ref.~\cite{ExtendedTwistedMass:2021gbo}.
\begin{table}[ht!]
    \centering
    \begin{tabular}{c|c|c|c|c|c}
       Ensemble  & $V/a^4$ & $\beta$ & a [fm] & $m_\pi L$ & \# meas.\\
       \hline 
        cB211.072.64 & $64^3 \times 128$ & 1.778 &  0.07975(32) & 3.62 &  48,000 \\
        cC211.06.80 & $80^3 \times 160$ & 1.836 &  0.06860(20) & 3.78 & 46,516\\
        cD211.054.96 & $96^3 \times 192$ & 1.900 &  0.05686(27) & 3.90 & 31,744
    \end{tabular}
    \caption{The parameters of the three $N_f=2+1+1$  ensembles used in this work. In the first column we give the name of the ensemble, in the second column the lattice volume,  in the third $\beta=6/g^2$  where $g$ the bare coupling constant, in the fourth the lattice spacing determined as discussed in Ref.~\cite{ExtendedTwistedMass:2021gbo}, and in the fifth column the value of $m_\pi L$. The last column is the number of measurements in the calculation of the three-point functions for $t_s/a=20$.}
    \label{tab:ens}
\end{table}

To evaluate the nucleon matrix elements of the operators in Eqs.~\eqref{Eq:vector} - \eqref{Eq:tensor}, we compute three- and two-point correlation functions. Gaussian  
smeared point sources are employed~\cite{Gusken:1989qx} to improve the overlap with the nucleon state. The connected three-point functions are computed using sequential propagators
inverted through the sink, i.e. using the so-called~\emph{fixed-sink} method.  In this work we restrict ourselves to the flavor non-singlet isovector combination where disconnected contributions vanish in the continuum limit. Connected three-point functions are computed using several  time separations, $t_s$, between the creation and annihilation  nucleon interpolating operators, namely $t_s \in [0.64,1.6]$~fm for the \texttt{cB211.072.64}, $t_s \in [0.55,1.52]$~fm for the \texttt{cC211.06.80} and $t_s \in [0.46,1.15]$~fm for the \texttt{cD211.054.96} ensemble. 
This broad range of separations is necessary for a thorough investigation and elimination of excited state contribution. At constant statistics, the noise-to-signal ratio increases exponentially with $t_s$ and the increase is exacerbated at the physical point. We thus increase the number of measurements with increasing $t_s$ to compensate, yielding an approximately constant error for all $t_s$. The desired ground state matrix element is obtained by taking an appropriate  ratio of three- to two-point functions (see Refs.~\cite{Alexandrou:2020okk,Hagler:2003jd}), and analyzing its time dependence as explained below.

In general, the nucleon matrix elements of the operators in Eqs.~\eqref{Eq:vector} - \eqref{Eq:tensor} yield linear combinations of the GFFs in the non-forward limit depending on the insertion operator quantum numbers, the nucleon spin projection, and components of the momentum transfer. We follow a standard procedure, as described in Sec.~C of Ref.~\cite{Alexandrou:2020okk}, where we construct an overconstrained system of equations that is inverted through a Singular Value Decomposition (SVD) to obtain the individual GFFs.

\begin{widetext}

\begin{figure}[ht!]
    \centering
    \includegraphics[width=\linewidth]{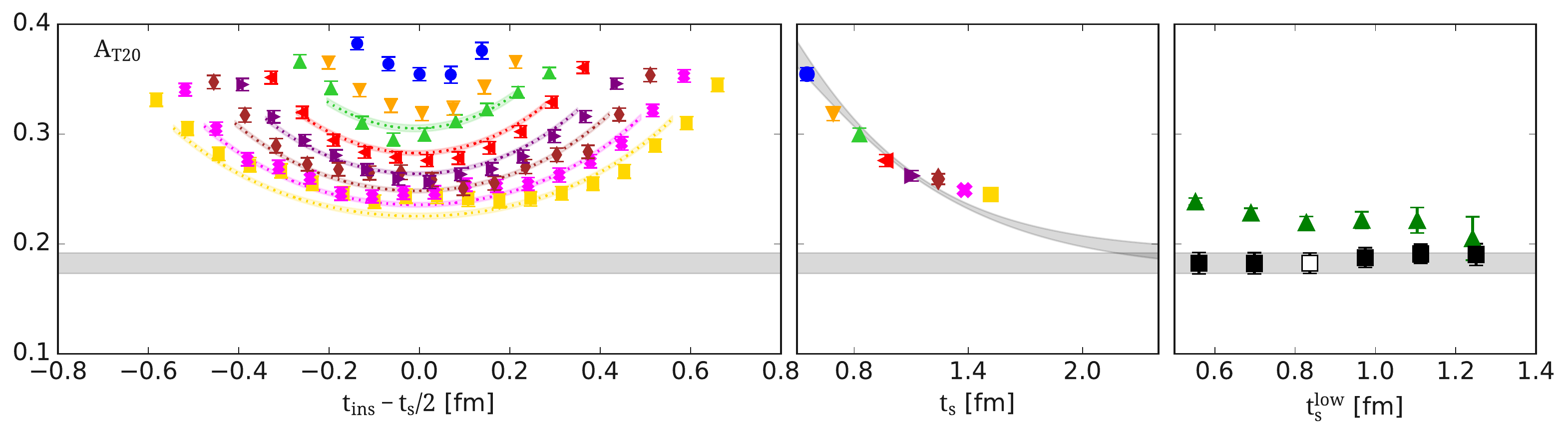}
    \caption{Excited states analysis for the determination of the matrix element  from which  $A_{T20}(0)$ is extracted for the \texttt{cC211.06.80} ensemble. In the left panel, we show the ratio of three- to two-point functions, for 
$t_s/a$ = 8, 10, 12, 14, 16, 18, 20, 22 with blue circles, orange down-pointing triangles, green up-pointing triangles, red left-pointing
triangles, purple right-pointing triangles, brown rhombuses, magenta crosses, and gold squares, respectively. The results are shown as a function
of the operator insertion time, $t_{\rm ins}$, shifted by $t_s/2$. In the middle panel, we show  the results extracted by fitting the ratios to a constant for each $t_s$ (plateau method), 
using the same symbols as in the left panel.  In the right panel we show the results from the 
summation (green triangles) and two-state (black squares) fits as we increase the smallest time separation, $t_s^{\rm low}$, used in
the fit. The open symbol and horizontal grey band spanning the three panels is the value we choose to determine the ground state matrix element. The parametric form of the two-state fit is used to predict the 
time-dependence of the ratio  shown with the grey curve in the middle panel and the colored bands shown in the left panel for each $t_s$.}
    \label{fig:ESA}
\end{figure}
\end{widetext}

A delicate step in our analysis is to esure that the ground state contribution is disentangled from the excited-states contamination. We follow the procedure of Ref.~\cite{Alexandrou:2020okk}, comparing three methods, namely, the plateau, summation, and  two-state fits. Both the plateau and summation fits take into account only contributions form the ground state, while in the two-state fit we consider contributions from the first excited state in both three- and two-point functions. An example analysis is shown in Fig.~\ref{fig:ESA} for the $A_{T20}(0)$ case. As can be seen, the ratio shows sizeable excited-states contamination. Including the first excited state in a two-state fit leads to a ground state matrix element that is significantly lower compared to the plateau method. For increasing $t_s^{\rm low}$ the summation fit agrees with the two-state fit, which is consistent for all $t_s^{\rm low}$. We therefore take the result of the two-state fit as the best determination of the ground state matrix element. This is done throughout our analysis of the GFFs.

The renormalization functions~\cite{Alexandrou:2019ali,Alexandrou:2010me,Alexandrou:2012mt} of the operators in Eqs.~\ref{Eq:vector} and \ref{Eq:tensor} are computed using the RI'-MOM~\cite{Martinelli:1994ty} scheme and results are converted to the ${\rm\overline{MS}}$ scheme at a scale of 4~GeV$^2$. A significant improvement in the determination of these renormalization functions comes from subtracting the
lattice artifact up to one-loop in perturbation theory~\cite{Alexandrou:2015sea}.

\begin{figure}[ht!]
    \centering
    \includegraphics[width=0.45\textwidth]{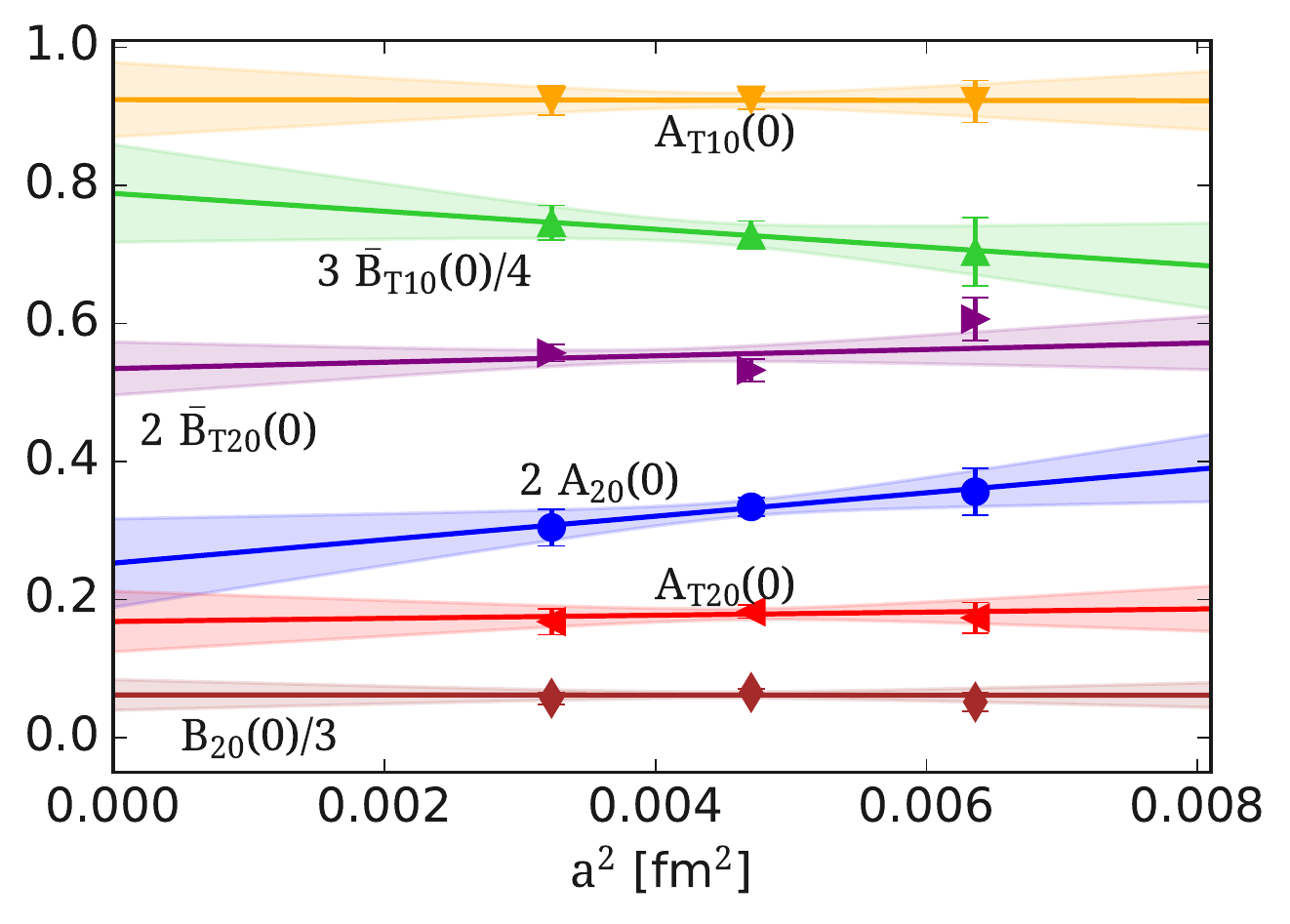}
    \caption{Continuum limit of selected unpolarized and tensor GFFs in the forward limit as a function of $a^2$. The lines with their associated error bands are linear fits in $a^2$. Results for $B_{20}(0)$, $A_{20}(0)$, $\bar{B}_{T20}(0)$, $\bar{B}_{T10}(0)$, $A_{T20}(0)$ and $A_{T10}(0)$ are presented with the name of each case being the closest to the corresponding band. We have scaled some of the quantities as indicated in the plot to avoid overlaps and improve presentation.
    Results are given in the ${\rm\overline{MS}}$ scheme at 4~GeV$^2$.}
    \label{fig:climQuant}
\end{figure}

\noindent
\textit{Results:}
\noindent 
In Fig.~\ref{fig:climQuant} we show the continuum limit of a selection of GFFs in the forward limit. Since our physical observables are automatically ${\cal O}(a)$-improved, we perform a linear fit in $a^2$ to extrapolate the results to $a \rightarrow 0$. As can be seen, for most of the cases the extrapolation is rather mild, which means that discretization effects are small for those quantities, within the current statistical precision.

In Table~\ref{tab:res}, we quote the values of the forward limit of the GFFs shown in Fig.~\ref{fig:climQuant} in the continuum limit. The quantity $g_T\equiv A_{T10}(0)$ is the tensor charge, which plays a crucial role  in the search of beyond the Standard Model (SM) interactions~\cite{Courtoy:2015haa} by experiments such as DUNE~\cite{Bischer:2018zcz} and IsoDAR~\cite{Abs:2015tbh}. Namely, the individual quark flavor contributions to $g_T$ enter  into the
determination of the quark electric dipole moment contribution to the neutron electric dipole moment~\cite{Bhattacharya:2015esa}, which if non-zero would signal the existence of physics beyond the SM. 
Determination of $g_T$ from phenomenology is achieved through the transversity PDF. Recent results using a global analysis of electron-proton
and proton-proton data have determined
$g_T=0.53(25)$~\cite{Radici:2018iag}. Although the central value is lower from our current determination, its error is  large, leading to about two standard deviations effect. Furthermore, our determination is fully compatible with the recent FLAG report~\cite{Aoki:2021kgd} and with our previous value~\cite{Alexandrou:2019brg} obtained using only the \texttt{cB211.072.64} ensemble, which is at the coarsest lattice spacing. 

Beyond $g_T$, another challenging quantity that is poorly known is the anomalous tensor magnetic moment $\kappa_T\equiv \bar{B}_{T10}(0)$.  It is a fundamental quantity, perhaps more  than $E_T$ and $\tilde{H}_T$~\cite{Diehl:2005jf},  describing the deformation of the transverse polarized quark
distribution in an unpolarized nucleon. First lattice results were presented in the pioneering work of the QCDSF/UKQCD collaboration~\cite{QCDSF:2006tkx}, where a value $\kappa_T=1.03(16)$ was reported obtained using chiral extrapolations from ensembles with pion masses of $m_\pi > 400$~MeV. Our analysis, using physical point ensembles, agrees with their value. Other results for this quantity include $\kappa_T=0.81$ and $1.24$ from two approaches using the constituent quark model~\cite{Pasquini:2005dk} and $\kappa_T=1.73$ using the quark-soliton model~\cite{Ledwig:2010zq}.
Since  $\kappa_T \sim - h_1^{\perp}$~\cite{Burkardt:2005hp}, then all results suggest that the Boer-Mulders function, $h_1^{\perp}$, should be negative and sizeable. This conclusion has also been found in a lattice QCD study of the transverse momentum dependent PDFs~\cite{Musch:2011er}. There, an $N_f=2+1$ mixed action scheme is used with domain wall valence fermions on Asqtad sea quarks and pion masses $m_\pi=369,518$~MeV.

\begin{table}[ht!]
    \centering
    \begin{adjustbox}{width=.5\textwidth,center}
    \begin{tabular}{|c|c|c|c|c|c|c|}
    \hline \hline
        $A_{T10}(0)$ & $\bar{B}_{T10}(0)$ & $A_{20}(0)$ & $B_{20}(0)$ & $J$ &$A_{T20}(0)$ & $\bar{B}_{T20}(0)$ \\
        \hline
        0.924(54) & 1.051(94) & 0.126(32) & 0.186(67) & 0.156(46) &  0.168(44) & 0.267(19) \\
        \hline 
    \end{tabular}
    \end{adjustbox}
    \caption{Our values of the forward limit of GFFs  presented in Fig.~\ref{fig:climQuant} in the continuum limit. We also include the value of the isovector light quark contribution to the nucleon angular momentum $(J)$.}
    \label{tab:res}
\end{table}

For the average momentum fraction, $\langle x \rangle \equiv A_{20}(0)$, our value is in agreement with the precise values from phenomenology~\cite{NNPDF:2017mvq,Dulat:2015mca,Harland-Lang:2014zoa}. While $\langle x \rangle$ is well-known, this is not the case for $B_{20}(0)$, which enters in the expression for the nucleon spin~\cite{Ji:1996ek}, $J=[A_{20}(0) + B_{20}(0)]/2$.  Having determined both GFFs  in the continuum limit we find  $J=0.156(46)$ for the isovector contribution which is compatible with our previous determination of 0.161(24)~\cite{Alexandrou:2019ali,Alexandrou:2020sml} obtained using only the \texttt{cB211.072.64} ensemble. The slightly larger value obtained here can be attributed to the slightly negative slope of $B_{20}(0)$ towards $a\rightarrow 0$ observed in Fig.~\ref{fig:climQuant}.

The second moment of the transversity PDF is $\langle x \rangle_{\delta u - \delta d} \equiv A_{T20}(0)$. Our finding is in agreement with our previous study using the \texttt{cB211.072.64} ensemble~\cite{Alexandrou:2019ali} and also with the value by the RQCD collaboration~\cite{Bali:2018zgl}.  $\bar{B}_{T20}(0)$ is  unknown from phenomenology. The lattice study by QCDSF/UKQCD~\cite{QCDSF:2006tkx}, using ensembles with pion masses $m_\pi>400$~MeV as discussed before, found $\bar{B}_{T20}(0)=0.160(39)$, which is about two standard deviations lower than our value.

\begin{figure}[ht!]
    \centering
    \includegraphics[width=0.45\textwidth]{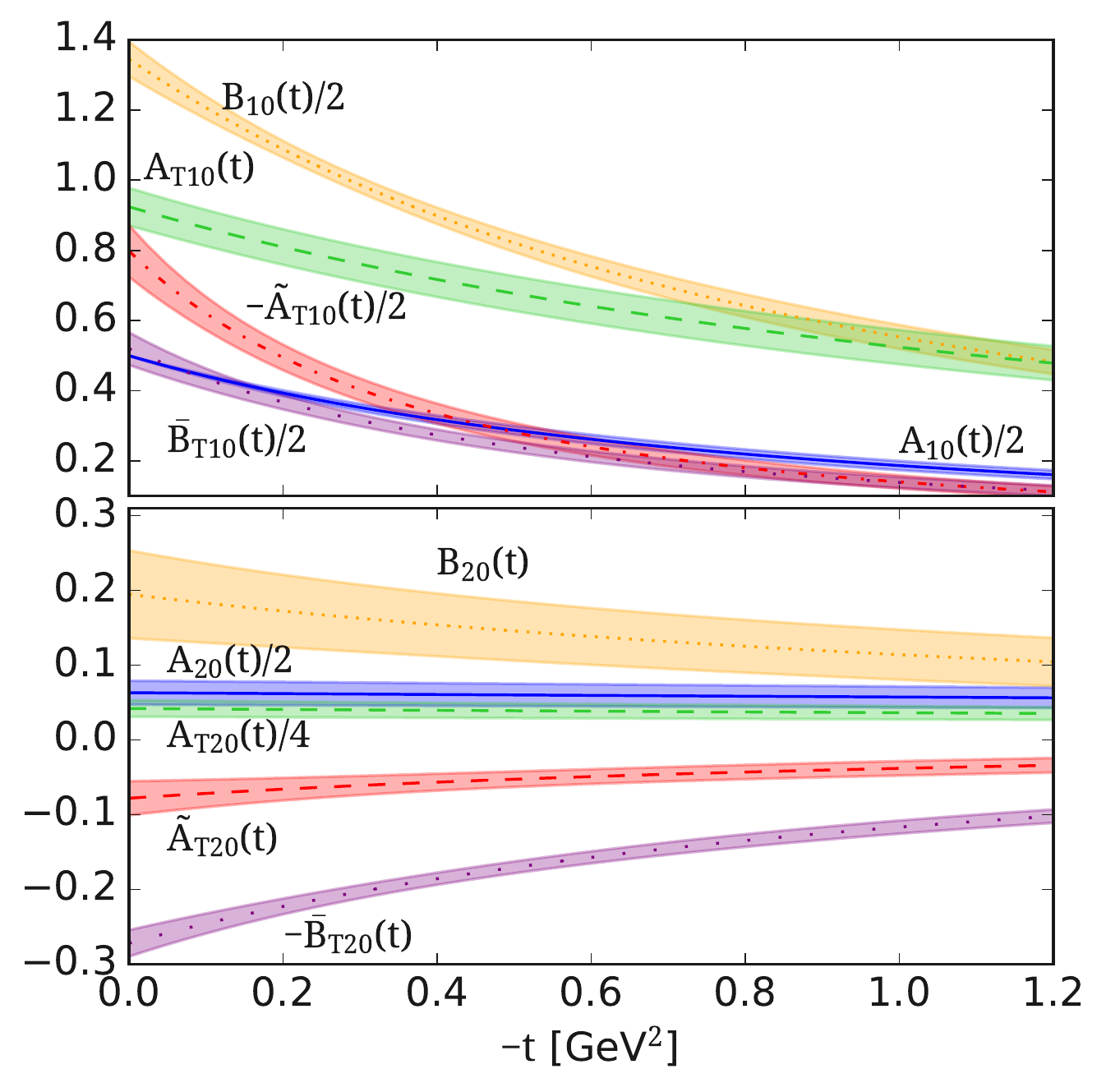}
    \caption{Results for  GFFs for $n=1$ (top) and $n=2$ (bottom) in the continuum limit as a function of the momentum transfer squared $-t=\Delta^2$. Results for $A_{n0}$, $B_{n0}$, $A_{Tn0}$, $\tilde{A}_{Tn0}$ and $\bar{B}_{Tn0}$ are presented with the name of each case being the closest to the corresponding band. We have scaled some GFFs as indicated in the plot to avoid overlaps and improve presentation.
    Results are given in the ${\rm\overline{MS}}$ at 2~GeV.}
    \label{fig:FFs}
\end{figure}

The dependence  of the GFFs on the momentum transfer squared, $-t$, is also extracted for each ensemble. Since in the lattice formulation  $-t$ takes  discrete values we  employ the $p$-pole Ansatz~\cite{QCDSFUKQCD:2006gmg,Diehl:2005jf},  
\begin{equation}
    F(t) = \frac{F(0)}{(1-t/m_p^2)^p},
    \label{Eq:ppole}
\end{equation}
to fit the GFFs. There are three fit parameters, namely $F(0)$, the value of the GFF in the forward limit, the pole mass $m_p$, and the value of $p$. Varying all three parameters leads to significant instabilities, as also observed in Refs.~\cite{QCDSF:2006tkx,Bali:2018zgl}. We use Gaussian priors for $p$ centered at $p=2$
with width 0.5. We find that this procedure leads to very stable results in all cases considered.   Note that for $A_{10}(t)$ and $B_{10}(t)$, i.e. the Dirac and Pauli form factors respectively, we use a dipole fit to parameterize their momentum dependence, therefore fixing $p=2$. 

In Fig.~\ref{fig:FFs} we show the vector and tensor GFFs in the continuum limit. With this information we can fully determine the first two moments of the transverse quark spin densities given in  Eq.~\eqref{Eq:rho}. As can be seen, the GFFs are well determined, especially for the $n=1$ case. As expected, for the higher moment, $n=2$, the GFFs have smaller values as compared to the $n=1$ GFFs. In addition, we observe that $A_{20}(t)$ and $A_{T20}(t)$ have a rather flat behavior. In impact parameter space, the fit function is given by~\cite{Diehl:2005jf}
\begin{eqnarray}
  F(b^2_\perp) = \frac{m_p^2 F(0)}{2^p \pi \Gamma(p)} (m_p b_\perp)^{p-1} K_{p-1}(m_p b_\perp),
\end{eqnarray}
where $\Gamma(x)$ is the Euler gamma function and $K_{n}(x)=K_{-n}(x)$ the modified Bessel functions and $b_\perp = \sqrt{b_\perp^2}$. 

\begin{figure}[ht!]
    \centering
    \includegraphics[width=0.49\textwidth]{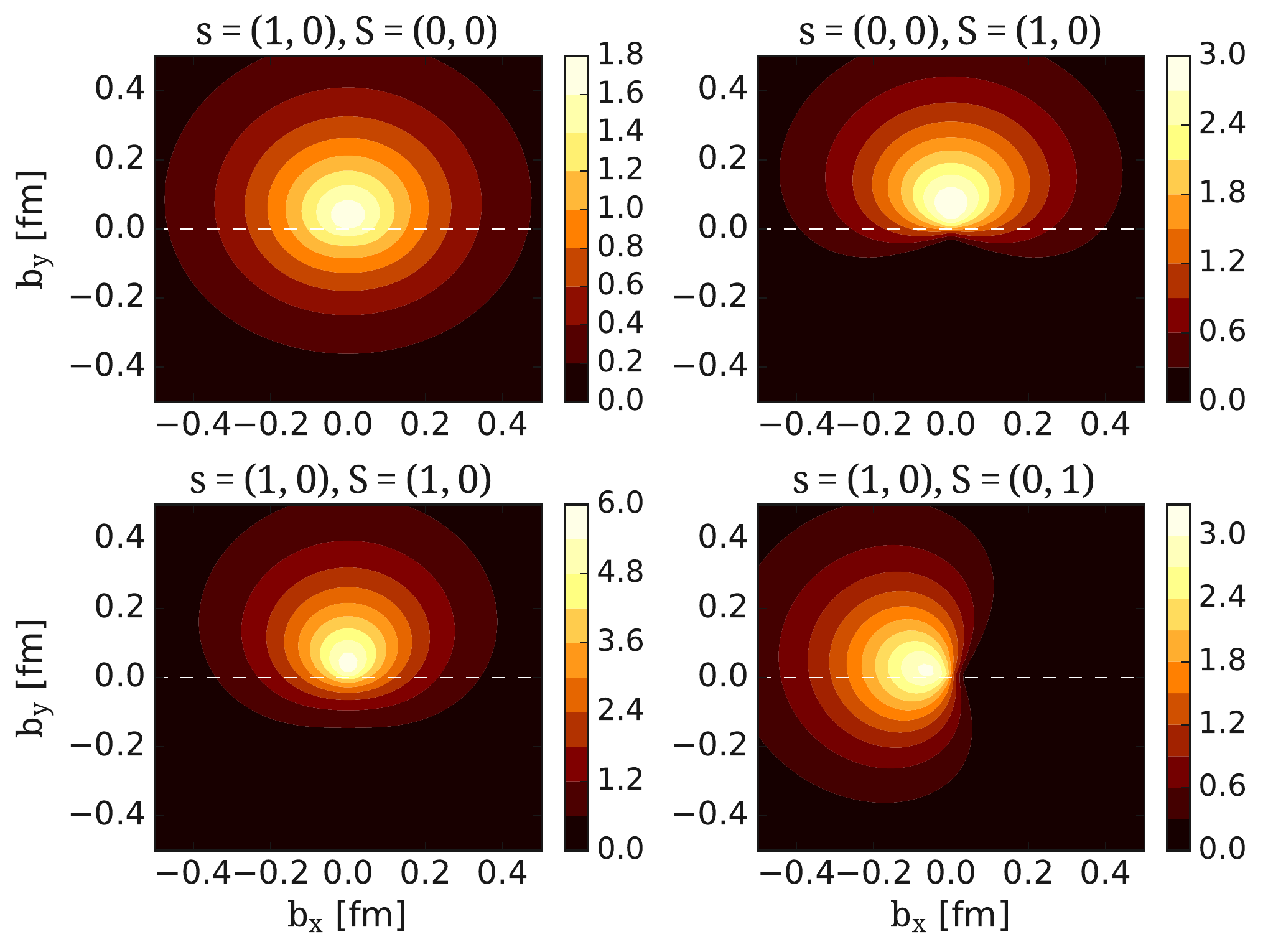}
    \caption{Contours of the first moment of the probability density defined in Eq.~\eqref{Eq:rho}, $\langle 1 \rangle_\rho$ [fm$^{-2}$] as a function of $b_x$ and $b_y$ in units of fm. Top-left: transversely polarized quarks in an unpolarized nucleon, top-right: unpolarized quarks in a transversely polarized nucleon, bottom-left: transversely polarized quarks in a transversely polarized nucleon and bottom-right: same as the bottom-left but with perpendicular polarizations between them. }
    \label{fig:x0}
\end{figure}

In Fig.~\ref{fig:x0} we show the first moment of the probability density $\rho(x,\bold{b}_\perp,\bold{s}_\perp,\bold{S}_\perp)$. It is very interesting that for all the cases we observe  a sizeable deformation. We consider four cases: i) For unpolarized quarks in a transversely polarized nucleon, we observe a huge distortion towards the positive $b_y$ direction. This can be traced back to the GFF $B_{10}$, contributing to the term for $E'$ in Eq.~\eqref{Eq:rho}, which from Fig.~\ref{fig:FFs} we see is large and drops fast yielding a large derivative. The origin of this behavior is related to the Sivers effect~\cite{Bury:2021sue}, a connection that has already been made in Ref.~\cite{Burkardt:2003uw}. ii) For transversely polarized quarks in an unpolarized nucleon, we can also observe a distortion, however, it is much milder compared to the previous case. This is because in the isovector combination the $\bar{B}_{T10}(b^2_\perp)$ term contributing here has a milder behavior compared to the individual quark behavior observed in Ref.~\cite{QCDSF:2006tkx}. iii) Another interesting case is when both quarks and the nucleon are transversely polarized. In this situation, all the terms in Eq.~\eqref{Eq:rho} contribute deforming significantly the density. iv) If one chooses the perpendicular polarization between the quarks and the nucleon, the third term drops out and the fourth one creates a significant impact, leading to a distortion also in the $b_x$ direction.

\begin{figure}[ht!]
    \centering
    \includegraphics[width=0.49\textwidth]{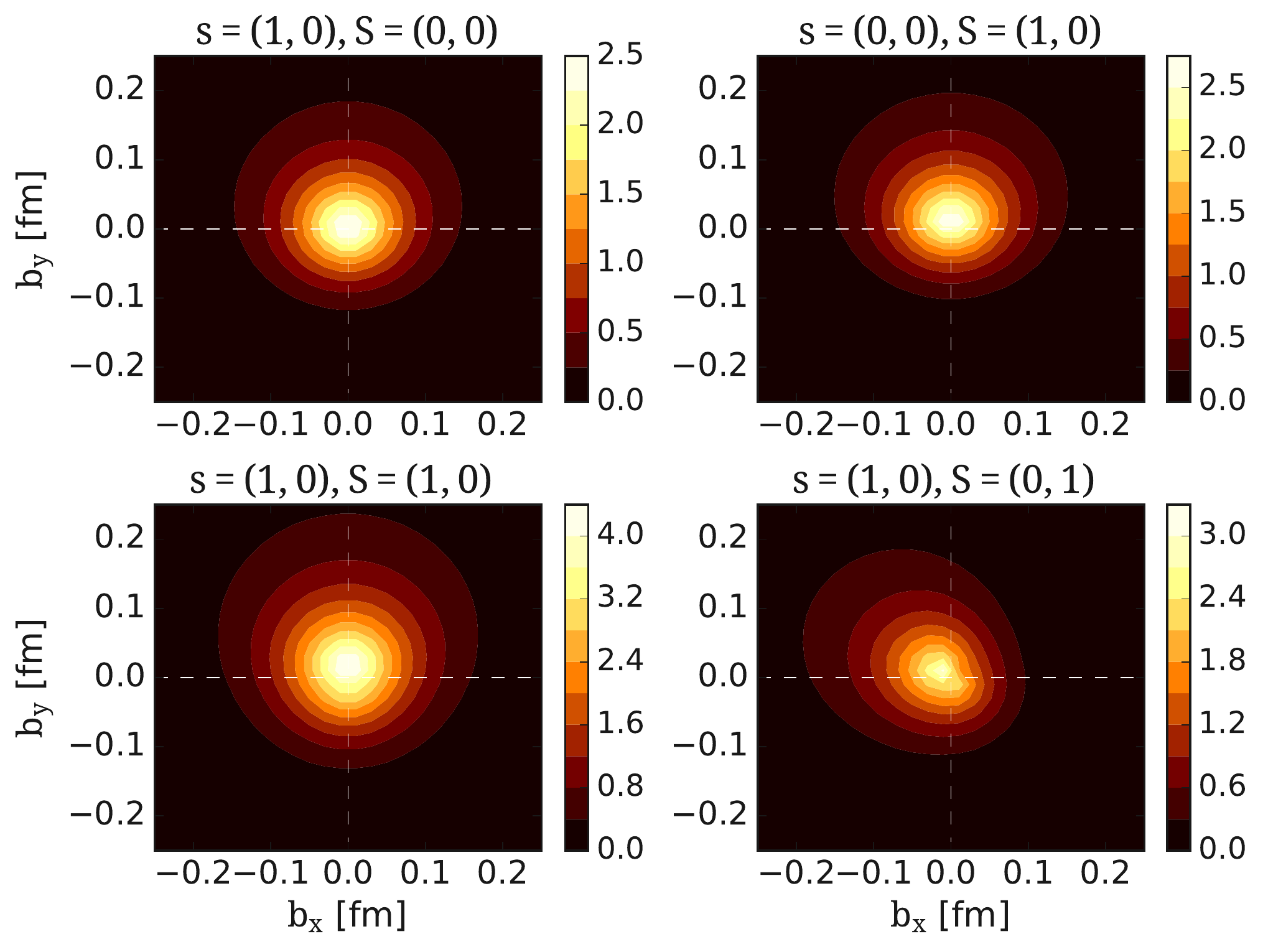}
    \caption{Contours of the second moment of the probability density defined in Eq.~\eqref{Eq:rho} $\langle x \rangle_\rho$ [fm$^{-2}$]. The notation is the same as in Fig.~\ref{fig:x0}.}
    \label{fig:x1}
\end{figure}

In Fig.~\ref{fig:x1} we show the second moment of the probability densities for the same four cases discussed in Fig.~\ref{fig:x0}. A general observation is that the distortion is milder and the densities are more localized around $\bold{b}_\perp=\bold{0}$. One reason is that $A_{20}(t)$ is relatively flat compared to  $A_{10}(t)$, leading to a rather localized density.  

\vspace*{0.25cm}
\noindent
\textit{Summary:}
\noindent
A lattice QCD calculation of the first two Mellin moments of the isovector transverse quark spin densities in the nucleon is presented. The calculation is performed using three twisted-mass fermion ensembles with lattice spacings  $a \simeq 0.057,\,0.069,\, 0.080~$fm enabling for the first time a controlled continuum extrapolation directly at the physical value of the pion mass. The extrapolation shows that discretization effects are mild for the targeted quantities. We confirm the existence of a sizeable Sivers and Boer-Mulders effect determining the anomalous tensor magnetic moment $\kappa_T=1.051(94)$. Results for the transverse quark spin densities demonstrate that significant deformations exist in the nucleon that are more prominent for the first moment. For the second moment the densities are more localized around the center of momentum of the proton. 

\vspace*{0.5cm}
\begin{acknowledgements}
We would like to thank all members of the Extended Twisted Mass Collaboration (ETMC) for a very constructive and enjoyable collaboration. M.C. acknowledges financial support by the U.S. Department of Energy Early Career Award under Grant No.\ DE-SC0020405. 
K.H. is financially supported by the Cyprus Research and Innovation foundation under contract number POST-DOC/0718/0100 and CULTURE-AWARD-YR/0220/0012. S.B., J.F. and K.H. are financially supported EuroCC project (GA No. 951732) funded by the Deputy Ministry of Research, Innovation and Digital Policy and the Cyprus Research and Innovation Foundation and
the European High-Performance Computing Joint Undertaking (JU) under grant agreement No 951732.
S.B. and J.F. are financially supported by the H2020 project PRACE 6-IP (GA No. 82376).
The project acknowledges support from the European Joint Doctorate projects HPC-LEAP and STIMULATE  funded by the European Union’s Horizon 2020 research and innovation programme under grant agreement No 642069 and 765048, respectively. G.S. acknowledges financial support from H2020 project PRACE-6IP (Grant agreement ID: 823767).
P.D. acknowledges support from the European Unions Horizon 2020 research and innovation programme under the Marie Sklodowska-Curie grant agreement No. 813942 (EuroPLEx) and  from INFN under the research project INFN-QCDLAT.
Results were obtained using Piz Daint at Centro Svizzero di Calcolo Scientifico (CSCS),
via the projects with ids s702, s954 and pr79. We thank the staff of CSCS for access to the computational resources and for their constant support.
The authors gratefully acknowledge the Gauss Centre for Supercomputing e.V. (www.gauss-centre.eu) for funding this project by providing computing time on the GCS Supercomputer JUWELS at Jülich Supercomputing Centre (JSC) through the projects FSSH, PR74YO and CECY00, CHCH02 (John von Neumann Institute for Computing (NIC)). Part of the results have been produced within the EA program of JUWELS Booster also with the help of the JUWELS Booster Project Team (JSC, Atos, ParTec, NVIDIA).
We acknowledge the CINECA award under the ISCRA initiative, for the availability of high performance computing resources and support. We acknowledge PRACE for awarding us access to Marconi100 at CINECA (Italy), Piz-Daint at CSCS (Switzerland) and Hawk at HLRS (Germany).
We acknowledge the Gauss Centre for Supercomputing e.V. (www. gauss-centre.eu) for project pr74yo by providing computing time on SuperMUC at LRZ (www.lrz.de) and Juwels Booster at JSC. The authors acknowledge the Texas Advanced Computing Center (TACC) at The University of Texas at Austin for providing HPC resources (Project ID PHY21001).
\end{acknowledgements}

\bibliography{refs}

\end{document}